# Model-Based Testing of Networked Applications

Yishuai Li
University of Pennsylvania
Philadelphia, PA, USA
yishuai@cis.upenn.edu

Benjamin C. Pierce
University of Pennsylvania
Philadelphia, PA, USA
bcpierce@cis.upenn.edu

Steve Zdancewic
University of Pennsylvania
Philadelphia, PA, USA
stevez@cis.upenn.edu

## ABSTRACT

We present a principled automatic testing framework for application-layer protocols. The key innovation is a domain-specific embedded language for writing nondeterministic models of the behavior of networked servers. These models are defined within the Coq interactive theorem prover, supporting a smooth transition from testing to formal verification.

Given a server model, we show how to automatically derive a tester that probes the server for unexpected behaviors. We address the uncertainties caused by both the server's internal choices and the network delaying messages nondeterministically. The derived tester accepts server implementations whose possible behaviors are a subset of those allowed by the nondeterministic model.

We demonstrate the effectiveness of this framework by using it to specify and test a fragment of the HTTP/1.1 protocol, showing that the automatically derived tester can capture RFC violations in buggy server implementations, including the latest versions of Apache and Nginx.

**ACM Reference Format:**
Yishuai Li, Benjamin C. Pierce, and Steve Zdancewic. 2021. Model-Based Testing of Networked Applications. In *Proceedings of the 30th ACM SIGSOFT International Symposium on Software Testing and Analysis (ISSTA '21), July 11–17, 2021, Virtual, Denmark.* ACM, New York, NY, USA, 11 pages. https://doi.org/10.1145/3460319.3464798

*This work is supported by the National Science Foundation's Expedition in Computing *The Science of Deep Specification*, award 1521539 (Weirich, Zdancewic, Pierce), with additional support by the NSF project *Random Testing for Language Design*, award 1421243 (Pierce). We thank Li-yao Xia for contributions to this work, and Yao Li, Hengchu Zhang, Jiani Huang, Yuepeng Wang, Qizhen Zhang, Yannan Li, Renqian Luo, Yushan Su, and Xiaojing Yu for comments on earlier drafts. We greatly appreciate the reviewers' comments and suggestions.

## 1 INTRODUCTION

The security and robustness of networked systems rest in large part on the correct behavior of various sorts of servers. This can be validated either by full-blown verification [4, 16] or model checking [2, 17, 19] against formal specifications, or (less expensively) by rigorous testing.

Rigorous testing requires a rigorous specification of the protocol that we expect the server to obey. Protocol specifications can be written as (1) a *server model* that describes *how* valid servers should handle messages, or (2) a *property* that defines *what* server behaviors are valid. From these specifications, we can conduct (1) *model-based testing* [7] or (2) *property-based testing* [12], respectively.

When testing server implementations against protocol specifications, one critical challenge is *nondeterminism*, which arises in two forms—we call them (1) *internal nondeterminism* and (2) *network nondeterminism*. (1) *Within* the server, correct behavior may be underspecified. For example, to handle HTTP conditional requests [10], a server generates strings called entity tags (ETags), but the RFC specification does not limit what values these ETags should be. Thus, to create test messages containing ETags, the tester must remember and reuse the ETags it has been given in previous messages from the server. (2) *Beyond* the server, messages and responses between the server and different clients might arrive in any order due to network delays and operating-system buffering. If the tester cannot control how the execution environment reorders messages—*e.g.,* when testing over the Internet—it needs to specify what servers are valid as observed over the network.

To address the challenges of both internal and network nondeterminism, we propose a generic framework for specifying and testing servers over the network. Our specification is phrased in terms of *interaction trees* [22], a general-purpose data structure for representing interactive processes. We specify the protocol with a server model (*i.e.*, a nondeterministic reference implementation), from which our framework automatically derives a tester that interacts with the server and tries to falsify the assertion that it meets its specification.

To handle internal nondeterminism in the server, we represent invisibly generated data and unknown initial state as symbolic expressions. These expressions are evaluated against observed messages during testing.

For network nondeterminism, we introduce a *network model* that describes how the network can delay messages. We compose this network model with the server model, yielding a model that exhibits all valid behaviors as observed by the client from across the network. Using this client-side model, our derived tester can interact with the server via multiple connections and reason about all possible message reorderings among the connections.

To validate our framework, we specified a fragment of HTTP/1.1, supporting WebDAV methods GET and PUT, conditional requests

(that are processed only if the precondition is satisfied), and proxying (forwarding the request to another HTTP server). The specification includes a server model for a subset of HTTP/1.1 and a network model for a subset of TCP. We derived a testing client from these models and found violations of *RFC 7232: HTTP/1.1 Conditional Requests* [10] in both Apache and Nginx. The tester was also able to capture several bugs that we intentionally inserted into Apache.

Our main contributions are:

- We propose a methodology for testing application layer protocols over the network. Our specification composes a server model with a generic network model, addressing network nondeterminism caused by network delays.
- We represent server states symbolically using interaction trees, allowing the testing framework to reason about internal nondeterminism in the server.
- We demonstrate the effectiveness of our methodology by specifying and testing a subset of HTTP/1.1 with both internal and network nondeterminism. Our automatically derived tester is able to detect violations of RFC standards in mainstream servers.

We describe the challenges of testing nondeterministic protocols in more detail in Section 2, define our specification language in Section 3, and explain our method for deriving testers from specifications in Section 4. We evaluate the derived tester for HTTP/1.1 empirically in Section 5, survey related work in Section 6, and conclude with future work in Section 7.

## 2 CHALLENGES: TESTING INTERNAL AND NETWORK NONDETERMINISM

To illustrate the challenges in testing networked applications, we discuss two features of HTTP/1.1—conditional requests [10] and message forwarding [11]—showcasing internal nondeterminism and network nondeterminism, respectively.

*Internal Nondeterminsm.* HTTP/1.1 requests can be conditional: if the client has a local copy of some resource and the copy on the server has not changed, then the server needn't resend the resource. To achieve this, an HTTP/1.1 server may generate a short string, called an "entity tag" (ETag), identifying the content of some resource, and send it to the client:

```
/* Client: */
GET /target HTTP/1.1

/* Server: */
HTTP/1.1 200 OK
ETag: "tag-foo"
... content of /target ...
```

The next time the client requests the same resource, it can include the ETag in the GET request, informing the server not to send the content if its ETag still matches:

```
/* Client: */
GET /target HTTP/1.1
If-None-Match: "tag-foo"

/* Server: */
HTTP/1.1 304 Not Modified
```

If the tag does not match, the server responds with code 200 and the updated content as usual. Similarly, if a client wants to modify the server's resource atomically by compare-and-swap, it can include the ETag in the PUT request as `If-Match` precondition, which instructs the server to only update the content if its current ETag matches.

Thus, whether a server's response should be judged *valid* or not depends on the ETag it generated when creating the resource. If the tester doesn't know the server's internal state (*e.g.*, before receiving any 200 response including the ETag), and cannot enumerate all of them (as ETags can be arbitrary strings), then it needs to maintain a space of all possible values, narrowing the space upon further interactions with the server.

It is possible, but tricky, to write an ad hoc tester for HTTP/1.1 by manually "dualizing" the behaviors described by the informal specification documents (RFCs). The protocol document describes *how* a valid server should handle requests, while the tester needs to determine *what* responses received from the server are valid. For example, "If the server has revealed some resource's ETag as `"foo"`, then it must not reject requests targetting this resource conditioned over `If-Match: "foo"`, until the resource has been modified"; and "Had the server previously rejected an `If-Match` request, it must reject the same request until its target has been modified." Figure 1 shows a hand-written tester for checking this bit of ETag functionality; we hope the reader will agree that this testing logic is not straightforward to derive from the informal "server's eye" specifications.

*Network Nondeterminism.* When testing an HTTP/1.1 server over the network, although TCP preserves message ordering within each connection, it does not guarantee any order between different connections. Consider a proxy model in Figure 2: it specifies how a server should forward messages. When the forwarded messages are scrambled as in Figure 3, the tester should be *loose* enough to accept the server, because a valid server may exhibit such reordering due to network delays. The tester should also be *strict* enough to reject a server that behaves as Figure 4, because no network delay can let the proxy forward a message before the observer sends it.

The kinds of nondeterminism exemplified here can be found in many other scenarios: (i) Servers may use some (unknown) algorithm to generate internal state for nonces, sequence numbers, caching metadata, *etc*, featuring internal nondeterminism. (ii) When the server runs multiple threads concurrently (*e.g.* to serve multiple clients), the operating system might schedule these threads nondeterministically. When testing the server over the network, such "nondeterminism outside the code of the server program but still within the machine on which the server is executing" is indistinguishable from nondeterminism caused by network delays, and thus can be covered by the concept "network nondeterminism."

## 3 SPECIFICATION LANGUAGE

A specification in our framework consists of two parts: a server model specifying server-side behavior, and a network model describing network delays. By composing these two models, we get a tester-side specification of valid observations over the network.

Formally, our specifications are written as *interaction trees*, a generic data structure for representing interactive programs in Coq.

```
(* update : (K → V) * K * V → (K → V) *)
let check (trace  : stream http_message,
           data   : key → value,
           is     : key → etag,
           is_not : key → list etag) =
  match trace with
  | PUT(k,t,v) :: SUCCESSFUL :: tr' ⇒
    if t ∈ is_not[k] then reject
    else if   is[k] == unknown
            ∨ strong_match(is[k],t)
         then let d' = update(data,k,v)     in
              let i' = update(is,k,unknown) in
              let n' = update(is_not,k,[])  in
         (* Now the tester knows that
          * the data in [k] is updated to [v],
          * but its new ETag is unknown. *)
              check(tr',d',i',n')
         else reject
  | PUT(k,t,v) :: PRECONDITION_FAILED :: tr' ⇒
    if strong_match(is[k],t) then reject
    else let n' = update(is_not, k, t::is_not[k])
      (* Now the tester knows that
       * the ETag of [k] is other than [t]. *)
         in check(tr',data,is,n')
  | GET(k,t) :: NOT_MODIFIED :: tr' ⇒
    if t ∈ is_not[k] then reject
    else if is[k] == unknown ∨ weak_match(is[k],t)
         then let i' = update(is,k,t) in
         (* Now the tester knows that
          * the ETag of [k] is equal to [t]. *)
              check(tr',data,i',is_not)
         else reject
  | GET(k,t0) :: OK(t,v) :: tr' ⇒
    if weak_match(is[k],t0) then reject
    else if data[k] ≠ unknown ∧ data[k] ≠ v
         then reject
         else let d' = update(data,k,v) in
              let i' = update(is,  k,t) in
         (* Now the tester knows
          * the data and ETag of [k]. *)
              check(tr',d',i',is_not)
  | _ :: _ :: _ ⇒ reject
  end
```

**Figure 1: Ad hoc tester for HTTP/1.1 conditional requests, for demonstrating how tricky it is to write the logic by hand. The checker determines whether a one-client-at-a-time `trace` is valid or not. The trace is represented as a stream (infinite linked list, constructed by "::") of HTTP messages sent and received. `PUT(k,t,v)` represents a PUT request that changes `k`'s value into `v` only if its ETag matches `t`; `GET(k,t)` is a GET request for `k`'s value only if its ETag does not match `t`; `OK(t,v)` indicates the request target's value is `v` and its ETag is `t`. The tester maintains three sorts of knowledge about the server: `data` stored for each content, what some ETag `is` known to be equal to, and what some ETag `is_not` equal to.**

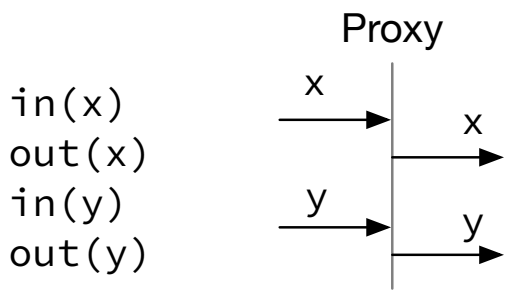

**Figure 2: Proxy model specifying a server that forwards a message immediately upon receiving it.**

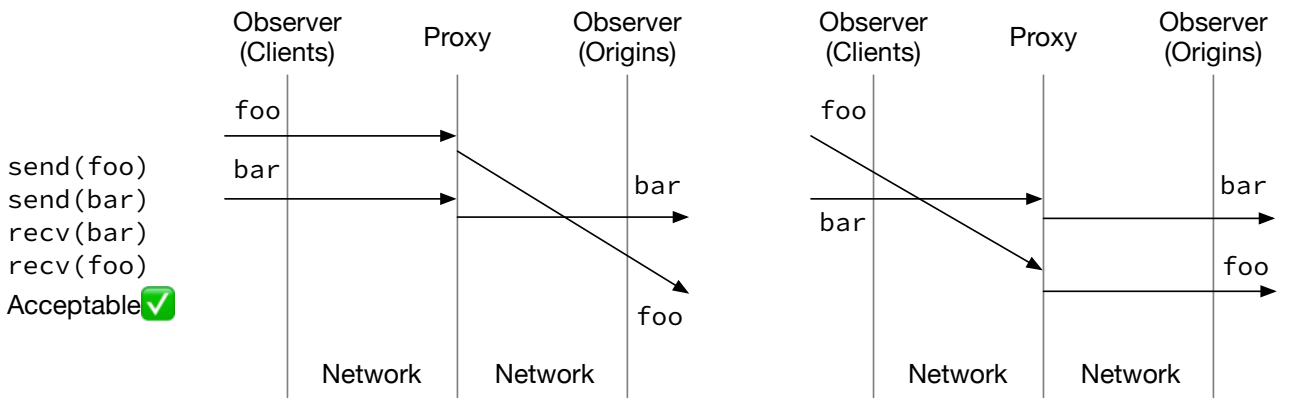

**Figure 3: A reordered observation, with two valid network-level explanations.**

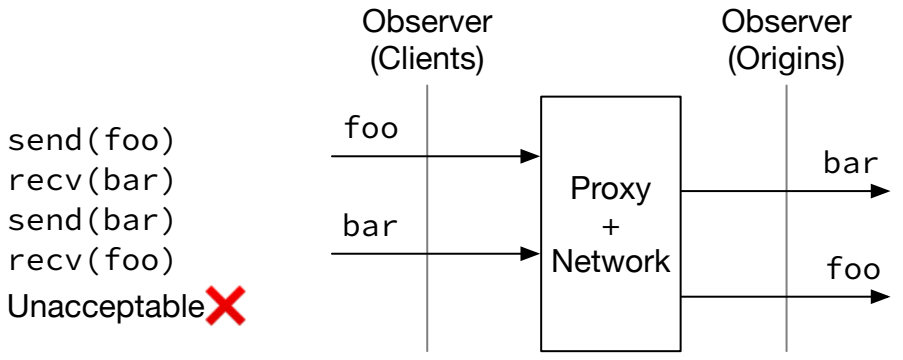

**Figure 4: Example of invalid observation.**

This language allows us to write rigorous mathematical specifications, and transform the specification into tester conveniently. In this paper, we present models as pseudocode for readability. Technical details about interaction trees can be found in [22].

Subsection 3.1 shows how to handle network nondeterminism. Subsection 3.2 then expands the model to address internal nondeterminism.

## 3.1 Server and Network Models

The *server model* specifies how the server code interacts with the network interface. For example, an extremely simplistic model of an HTTP proxy (shown in Figure 2) is written as:

```
let proxy() =
    msg := recv();
    send(msg);
    proxy()
```

An implementation is said to be *valid* if it is indistinguishable from the model when viewed from across the network. Consider the following proxy implementation that reorders messages:

```
void proxy_implementation() {
  while (true) {
    recv(&msg1); recv(&msg2);
    send(msg2);  send(msg1);
  }
}
```

```
let tcp (buffer : list packet) =
    let absorb =
        pkt := recv();
        tcp (buffer ++ [pkt]) in
    let emit =
        let pkts = oldest_in_each_conn(buffer) in
        pkt := pick_one(pkts);
        send(pkt);
        tcp (remove(pkt, buffer)) in
    or (absorb, emit)
```

**Figure 5: Network model for concurrent TCP connections. The model maintains a `buffer` of all packets en route. In each cycle, the model may nondeterministically branch to either absorb `or` emit a packet. Any absorbed packet is appended to the end of buffer. When emitting a packet, the model may choose a connection and send the oldest packet in it.**

This reordered implementation is valid, because the model itself may exhibit the same behavior when observed over the network, as shown in Figure 3. This "implementation's behavior is explainable by the model, considering network delays" relation is called *network refinement* by Koh et al. [15].

To specify network refinement in a testable way, we introduce the *network model*, a conceptual implementation of the transport-layer environment between the server and the tester. It models the network as a nondeterministic machine that absorbs packets and, after some time, emits them again. Figure 5 shows the network model for concurrent TCP connections: The network either receives a packet from some node, or sends the first packet en route of some connection. This model preserves the message order within each connection, but it exhibits all possible reorderings among different connections.

The network model does not distinguish between server and tester. When one end `send`s some message, the network `recv`s the message and `send`s it after some cycles of delay; it is then observed by the other end via some `recv` call.

In Subsection 4.3, we compose the server and network models to yield an observer-side specification for testing purposes.

## 3.2 Symbolic Representation of Nondeterministic Data

To incorporate symbolic evaluation in our testing framework, our specification needs to represent internally generated data as symbols. Consider HTTP PUT requests with `If-Match` preconditions: Upon success, the server generates a new ETag for the updated content, and the tester does not know the ETag's value immediately. Our symbolic model in Figure 6 represents the server's generated ETags as fresh variables. The server's future behavior might depend on whether a request's ETag matches the generated (symbolic) ETag. Such matching produces a symbolic boolean expression, which cannot be evaluated into a boolean value without enough constraints on its variables. Our model introduces `IF` operator to condition branches over a symbolic boolean expression. Which branch the server actually took is decided by the derived tester in Section 4.

```
(* matches : (etag * exp etag) → exp bool *)
(* IF      : (exp bool * T * T) → T       *)
let put (k    : key,
         t    : etag,
         v    : value,
         data : key → value,
         xtag : key → exp etag) =
    IF (matches(t, xtag[k]),
    (* then *)
       xt := fresh_tag();
       let xtag' = update(xtag, k, xt) in
       let data' = update(data, k, v)  in
       return (OK, xtag', data'),
    (* else *)
       return (PreconditionFailed, xtag, data))
```

**Figure 6: Symbolic model handling conditional PUT request. The model maintains two states: `data` that maps keys to their values, and `xtag` that maps keys to symbolic variables that represent their corresponding ETags. Upon receiving a PUT request conditioned over "If-Match: `t`", the server should decide whether the request ETag `matches` that stored in the server. Upon matching, the server processes the PUT request, and represents the updated value's ETag as a fresh variable.**

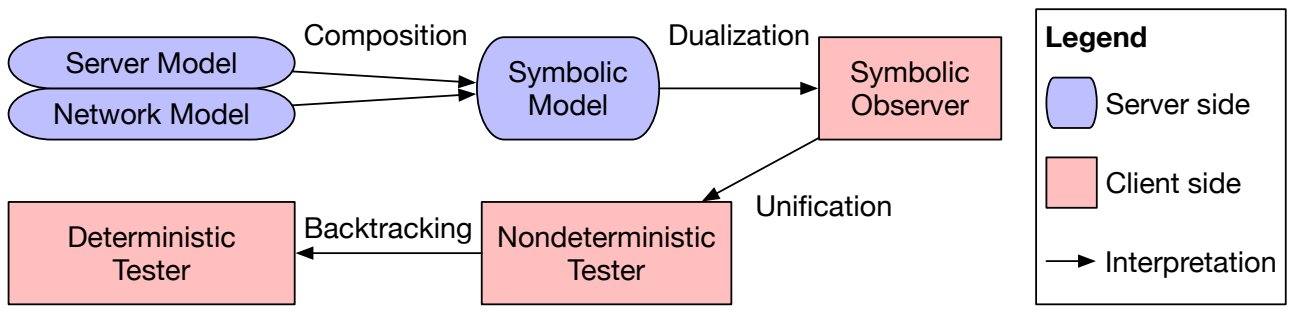

**Figure 7: Deriving tester program from specification**

In Subsection 4.2, we implement the symbolic evaluation process that checks servers' observable behavior against this symbolic model.

# 4 DERIVATION: FROM SERVER SPECIFICATION TO TESTING PROGRAM

From the specified the application and network models, our framework automatically derives a tester program that interacts with the server and determines its validity. The derivation framework is shown in outline in Figure 7. Each box is an interaction tree program, and the arrows are "interpretors" that transform one interaction tree into another. Subsection 4.1 explains the concept of interpretation, and the rest of this section describes how to interpret the specification into a tester program.

## 4.1 Interpreting Interaction Trees

Interaction tree programs can be destructed into an interaction event followed by another interaction tree program. Such structure allows us to *interpret* one program into another. Figure 8 shows an example of interpretation: The original `acc` program sends and receives messages, and the `tee` interpretor transforms the `acc` into another program that also prints the messages sent and received.

```
let acc(sum) =
  x := recv(); send(x+sum); acc(x+sum) in
let tee(m) =
  match m with
  | x := recv(); m'(x)⇒
    a := recv(); print("IN" ++ a); tee(m'(a))
  | send(a); m'⇒
    print("OUT" ++ a); send(a); tee(m')
  end in
tee(acc(0))
(* ... is equivalent to ... *)
let tee_acc(sum) =
  a := recv(); print("IN" ++ a);
  print("OUT" ++ (a+sum)); send(a+sum);
  tee_acc(a+sum) in
tee_acc(0)
```

**Figure 8: Interpretation example. `acc` receives a number and returns the sum of numbers received so far. `tee` prints all the numbers sent and received. Interpreting `acc` with interpretor `tee` results in a program that's equivalent to `tee_acc`.**

```
1  let observe (server) =
2    match server with
3    | pkt := recv(); s'(pkt)⇒
4      p := gen_pkt(); send(p); observe (s'(p))
5    | send(pkt); s'⇒
6      p := recv(); guard(pkt, p); observe (s')
7    | IF (x, s1, s2)⇒
8      (* Allow validating observation with [s1],
9       * provided [x] is unifiable with [true];
10      * Or, unify [x] with [false],
11      * and validate observation with [s2]. *)
12     determine(unify(x, true ); observe (s1),
13               unify(x, false); observe (s2))
14   | r  := _(); s'(r)⇒
15     r1 := _(); observe (s'(r1))
16   end
```

**Figure 9: Dualizing server model into observer model. Upon `recv` events, the observer generates a packet and sends it to the server. For `send` events, the observer receives a packet `p1`, and fails if it does not match the specified `pkt`. When the server makes nondeterminstic `IF` branches, the observer `determine`s between the branches by `unify`ing the branch condition with its conjectured value, and then observing the corresponding branch.**

Such interpretation is done by pattern matching on the program's structure in Line 4. Based on what the original program wants to do next, the interpretor defines what the result program should do in Line 6 and Line 8. These programs defined in accordance to events are called *handlers*. By writing different handlers for the events, interpretors can construct new programs in various ways, as shown in following subsections. Further details about interpreting interaction trees are explained by Xia et al. [22].

## 4.2 From Server Specification to Tester Program

For simplicity, we first explain how to handle servers' internal nondeterminism with symbolic evaluation. This subsection covers a subgraph of Figure 7, starting with dualizing the symbolic model. Here we use the server model itself as the symbolic model, assuming no reorderings by network delays. We will compose the server model with the network model in Subsection 4.3, addressing network nondeterminism.

*Dualization.* To *observe* the server's behavior, we have to interpret the specified server-side events into tester-side events: When the server should send a certain message, the tester expects to receive the specified message, and rejects the server upon receiving an unexpected message; when the server should receive some message, the tester generates a message and sends it to the server, as shown in Figure 9.

Besides sending and receiving messages, the model also has `IF` branches conditioned over symbolic expressions, like that shown in Figure 6. Upon nondeterministic branching, the tester needs to determine which branch was actually taken, by constructing observers for both branches. Each branch represents a possible explanation of the server's behavior. Upon further interacting with the server, some branches might fail because its conjecture cannot explain what it has observed. The tester rejects the server if all branches have failed, indicating that the server corresponds to no possible case in the model.

Dualizing the server-side model produces an observer model that performs interactions to reveal the server's behavior and check its validity. This model includes all possible observations from a valid server, and needs to `determine` which branch in the server model matches the observed behavior. The model validates its observations with unification events `unify` and `guard`. These primitive events are handled by later interpretations: The `unify` and `guard` events in each branch are instantiated into symbolic evaluation logic that decides whether this branch should fail or not; The `determine` events are instantiated into backtracking searches to find if all branches have failed, which rejects the server.

*Symbolic Evaluation.* In this interpretation phase, we handle nondeterminism at data level by handling `fresh` events in the server model, as well as `unify` and `guard` events introduced by dualization. The interpretor instantiates these events into symbolic evaluation algorithms.

As shown in Figure 10 (skip Line 18–28 for now—we'll explain that part later), the tester checks whether the observed/conjectured value matches the specification, by maintaining the constraints on the symbolic variables. These constraints are initially empty when the variables are generated by `fresh` events. As the test runs into `unify` and `guard` events, it adds constraints `assert`ing that the observed value matches the specification, and checks whether the constraints are still compatible. Incompatibility among constraints indicates that the server has exhibited behavior that cannot be explained by the model, implying violation against the current branch of specification.

```
1 (* unifyS = list variable * list constraint    *)
2 (* new_var : unifyS → variable * unifyS        *)
3 (* assert : exp T * T * unifyS → option unifyS *)
4 let unifier (observer, map : mcid → pcid,
5          vars : unifyS) =
6   match observer with
7   | x := fresh(); o'(x)⇒
8     let (x1, vars') = new_var(vars) in
9     unifier (o'(x1), vars', map)
10  | unify(x, v); o'⇒
11    match assert(x, v, vars) with
12    | Some vars'⇒ unifier (o', vars', map)
13    | None⇒ failwith "Unexpected payload"
14    end
15  | guard(p0, p1); o'⇒
16    match assert(p0, p1, vars) with
17    | Some vars'⇒
18      let mc = p0.source in
19      let pc = p1.source in
20      if mc.is_created_by_server
21      then match map[mc] with
22           | pc⇒ unifier (o', vars', map)
23           | unknown⇒
24             let map' = update(map, mc, pc) in
25             unifier (o', vars', map')
26           | others⇒
27             failwith "Unexpected connection"
28           end
29      else unifier (o', vars', map)
30    | None⇒ failwith "Unexpected payload"
31    end
32  | r  := _(); o'(r)⇒
33    r1 := _(); unifier (o'(r1), vars, map)
34  end
```

**Figure 10: Instantiating symbolic events. The tester maintains a unifyState which stores the constraints on symbolic variables. When the specification creates a fresh symbol, the tester creates an entry for the symbol with no initial constraints. Upon unify and guard events, the tester checks whether the assertion is compatible with the current constraints. If yes, it updates the constraints and move on; otherwise, it raises an error on the current branch.**

```
(* filter : event T * T * list M → list M *)
(* [filter(e, r, l)] returns a subset in [l],
 * where the model programs' next event is [e]
 * that returns [r]. *)
let backtrack (current, others) =
  match current with
  | determine(t1, t2)⇒
    backtrack (t1, t2::others)
  | failwith error⇒ (* current branch failed *)
    match others with
    | []⇒ failwith error
    | another::ot'⇒ backtrack (another, ot')
    end
  | send(pkt); t'⇒
    let ot' = filter(SEND, pkt, others) in
    send(pkt); backtrack (t', ot')
  | pkt := recv(); t'(pkt)⇒
    opkt := maybe_recv();
    match opkt with
    | Some p1⇒
      let ot' = filter(RECV, pkt, others) in
      backtrack (t'(p1), ot')
    | None⇒             (* no packet arrived *)
      match others with
      | []⇒ backtrack (current, []) (* retry *)
      | another::ot'⇒             (* postpone *)
        backtrack (another, ot'++[current])
      end
    end
  end in
backtrack (tester_nondet, [])
```

**Figure 11: From nondeterministic model to deterministic tester program. If the model makes nondeterministic branches, the tester picks a branch to start with, and puts the other branch into a set of other possibilities. If the current branch has failed, the tester looks for other possible branches to continue checking. When the current branch sends a packet, the tester filters the set of other possibilities, and only keeps the branches that match the current send event. If the model wants to receive a packet, the tester handles both cases whether some packet has arrived or not.**

*Handling Incoming Connections.* In addition to generating data internally, the server might exhibit another kind of nondeterminism related to the outgoing connections it creates. For example, when a client uses an HTTP server as proxy, requesting resources from another server, the proxy server should create a new connection to the target server. However, as shown in Figure 3, when the tester receives a request from an accepted connection, it does not know which client's request the proxy was forwarding, due to network delays.

Outgoing connections created by the server model are identified by "model connection identifiers" (mcid), and the tester accepts incoming connections identified by "physical connection identifiers" (pcid). As shown in Line 18–28 of Figure 10, to determine which mcid in the specification does a runtime pcid corresponds to, the tester maintains a mapping between the connection identifiers. Such mapping ensures the tester to check interactions on an accepted connection against the right connection specified by the server model.

*Backtracking.* Symbolic evaluation determines whether the observations matches the tester's conjectures on each branch. So far, the derived tester is a nondeterministic program that rejects the server if and only if all possible branches have raised some error. To simulate this tester on a deterministic machine, we execute

```
let http_server (http_st) =
  request := recv_HTTP(http_st);
  (response, st') := process(request, http_st);
  http_server (st')
...
let observer (server) =
  match server with
  | req := recv_HTTP(http_st); s'(req)⇒
    r1  := gen_Observer(http_st);
    send(r1); observe (s'(r1))
...
let unifier (observer, vars, conn) =
  match observer with
  | req := gen_Observer(http_st); o'(req)⇒
    r1  := gen_Unifier(http_st, vars, conn);
    unifier (o'(r1), vars, conn)
...
```

**Figure 12: Embedding programs' internal state into the events. By expanding the events' parameters, we enrich the test case generator's knowledge along the interpretations.**

one branch until it fails. Upon failure in the current branch, the simulator switches to another possible branch, until it exhausts all possibilities and rejects the server, as shown in Line 9–13 of Figure 11.

When switching from one branch to another, the tester cannot revert its previous interactions with the server. Therefore, it must match the server model against all interactions it has performed, and filter out the mismatching branches, as shown in Line 15 and Line 21 of Figure 11.

We've now derived a tester from the server model. The specified server runs forever, and so does the tester (upon no violations observed). We accept the server if the tester hasn't rejected it after some large, pre-determined number of steps of execution.

*Test Case Generation.* Counterexamples are sparsely distributed, especially when the bugs are related to server's internally generated data like ETags, which can hardly be matched by a random test case generator. After observing the `ETag` field of some response, the generator can send more requests with the same ETag value, rather than choosing an unknown value arbitrarily.

As shown in Figure 12, our derivation framework allows passing the programs' internal state as the events' parameters, so the test case generator can utilize the states in all intermediate interpretation phases, and apply heuristics to emphasise certain bug patterns.

Notice that the state-passing strategy only allows tuning *what* messages to send. To reveal bugs more efficiently in an interactive scenario, we need to tune *when* the interactions are made, which is further discussed in Subsection 5.2. Generating test cases in certain orders is to be explored in future work.

```
1  let compose (net, bi, bo, srv) =
2    let step_net =
3      match net with
4      | send(pkt); n'⇒
5        if pkt.to_server
6        then compose (n', bi++[pkt], bo, srv)
7        else send(pkt);   (* to client *)
8             compose (n', bi, bo, srv)
9        end
10     | pkt := recv(); n'(pkt)⇒
11       match bo with
12       | p0::b'⇒ compose (n'(p0), bi, b', srv)
13       | []    ⇒ p1 := recv();
14                 compose (n'(p1), bi, bo, srv)
15       end
16     | r  := _(); n'(r)⇒
17       r1 := _(); compose (n'(r1), bi, bo, srv)
18     end in
19   match srv with
20   | send(pkt); s'⇒
21     compose (net, bi, bo++[pkt], s')
22   | pkt := recv(); s'(pkt)⇒
23     match bi with
24     | p0::b'⇒ compose (net, b', bo, s'(p0))
25     | []    ⇒ step_net
26     end
27   | r  := _(); s'(r)⇒
28     r1 := _(); compose (net, bi, bo, s'(r1))
29   end in
30 compose (tcp, [], [], http)
```

**Figure 13: Composing `http` server model with `tcp` network model by interpreting their events and passing messages from one model to another. The composing function takes four parameters: server and network models as `srv` and `net`, and the message buffers between them. When `srv` wants to `send` a packet in Line 21, the packet is appended to the outgoing buffer `bo` until absorbed by `net` in Line 12, and eventually emitted to the client in Line 7. Conversely, packets sent by clients are absorbed by `net` in Line 13, emitted to the application's incoming buffer `bi` in Line 6, until `srv` consumes it in Line 24.**

## 4.3 Network Composition

We have shown how to derive a tester from the server model itself. The server model describes how a reference server processes messages. For protocols like HTTP/1.1 where servers are expected to handle one request at a time, a reasonable server model should be "linear" that serves one client after another. As a result, the derived tester only simulates a single client, and does not attempt to observe the server's behavior via multiple simultaneous connections.

The network model describes how messages sent by one end of the network are eventually received by the other end. When interacting with multiple clients, a valid server's observable behavior should be explainable by "server delayed by the network", as discussed in Subsection 3.1. To model this set of observations, we compose the server and network models by attaching the server model as one end on the network model.

As shown in Figure 13, we `compose` the events of server and network models. Messages sent by the server are received by the network and sent to clients after some delay, and vice versa. Such composition produces a model that branches nondeterministically, and includes all possible interactions of a valid HTTP server that appear on the client side.

The composed model does not introduce new events that were not included in the server model: The network model in Figure 5 does perform nondeterministc `or` branches, but `or(x,y)` is a syntactic sugar for `b := fresh(); IF(b,x,y)`. Therefore, using the same derivation algorithm from the server model to single-connection tester program, we can derive the composed server+network model into a multi-connection tester.

Notice that the server and network events are scheduled at different priorities: The composition algorithm steps into the network model lazily, not until the server is blocked in Line 25. When the network wants to `recv` some packet in Line 10, it prioritizes packets sent by the server, and only receives from the clients if the server's outgoing buffer has been exhausted. Such design is to enforce the tester to terminate upon observing invalid behavior: When the server's behavior violates the model, the tester should check all possible branches and determine that none of them can lead to such behavior. If the model steps further into the network, it would include infinitely many `absorb` branches in Figure 5, so the derived tester will never exhaust "all" branches and reject the server. Scheduling network events only when the server model is blocked produces sufficient nondeterminism to accept valid servers.

# 5 EVALUATION

To evaluate whether our derived tester is effective at finding bugs, we ran the tester against mainstream HTTP servers, as well as server implementations with bugs inserted by us.

## 5.1 Experiment Setup

*Systems Under Test (SUTs).* We ran the tests against Apache HTTP Server [9], which is among the most popular servers on the World Wide Web. We used the latest release 2.4.46, and edited the configuration file to enable WebDAV and proxy modules. Our tester found a violation against RFC 7232 in the Apache server, so we modified its source code before creating mutants.

We've also tried testing Nginx and found another violation against RFC 7232. However, the module structure of Nginx made it difficult to fix the bug instantly. (The issue was first reported 8 years ago and still not fixed!) Therefore, no mutation testing was performed on Nginx.

*Infrastructure.* The tests were performed on a laptop computer (with Intel Core i7 CPU at 3.1 GHz, 16GB LPDDR3 memory at 2133MHz, and macOS 10.15.7). The SUT was deployed as a Docker instance, using the same host machine as the tester runs on. They communicate with POSIX system calls, in the same way as over Internet except using address `localhost`. The round-trip time (RTT) of local loopback is 0.08 ± 0.04 microsecond (at 90% confidence).

## 5.2 Results

*Finding Bugs in Real-World Servers and Mutants.* Our tester rejected the unmodified Apache HTTP Server, which uses strong comparison for PUT requests conditioned over `If-None-Match`, while RFC 7232 specified that `If-None-Match` preconditions must be evaluated with weak comparison. We reported this bug to the developers, and figured out that Apache was conforming with an obsoleted HTTP/1.1 standard [8]. The latest standard has changed the semantics of `If-None-Match` preconditions, but Apache didn't update the logic correspondingly.

We created 20 mutants by manually modifying the Apache source code. The tester rejected all the 20 mutants, located in various modules of the Apache server: `core`, `http`, `dav`, and `proxy`. They appear both in control flow (*e.g.*, early return, skipped condition) and in data values (*e.g.*, wrong arguments, flip bit, buffer off by one byte).

We didn't use automatic mutant generators because (i) Existing tools could not mutate all modules we're interested in; and (ii) The automatically generated mutants could not cause semantic violations against our protocol specification.

When testing Nginx, we found that the server did not check the preconditions of PUT requests. We then browsed the Nginx bug tracker and found a similar ticket opened by Haverbeke [13]. These results show that our tester is capable of finding bugs in server implementations, including those we're unaware of.

*Performance.* As shown in Figure 14, the tester rejected all buggy implementations within 1 minute. In most cases, the tester could find the bug within 1 second.

Some bugs took longer time to find, and they usually required more interactions to reveal. This may be caused by (1) The counterexample has a certain pattern that our generator didn't optimize for, or (2) The tester did produce a counter-example, but failed to reject the wrong behavior. We determine the real cause by analysing the bugs and their counterexamples:

- Mutants 19 and 20 are related to the WebDAV module, which handles PUT requests that modify the target's contents. The buggy servers wrote to a different target from that requested, but responds a successful status to the client. The tester cannot tell that the server is faulty until it queries the target's latest contents and observes an unexpected value. To reject the server with full confidence, these observations must be made in a certain order, as shown in Figure 15.
- Mutant 18 is similar to the bug in vanilla Apache: the server should have responded with 304 Not Modified, but sent back 200 OK instead. To reveal such violation, a minimal counterexample consists of 4 messages: (1) GET request, (2) 200 OK response with some ETag `x`, (3) GET request conditioned over `If-None-Match: x`, and (4) 200 OK response, indicating that the ETag `x` did not match itself. Notice that (2) must be observed before (3), otherwise the tester will not reject the server, with a similar reason as Figure 15.
- Mutant 5 causes the server to skip some code in the core module, and send nonscence messages when it should respond with 404 Not Found. The counterexample can be as small as one GET request on a non-existential target, followed by a non-404, non-200 response. However, our tester generates

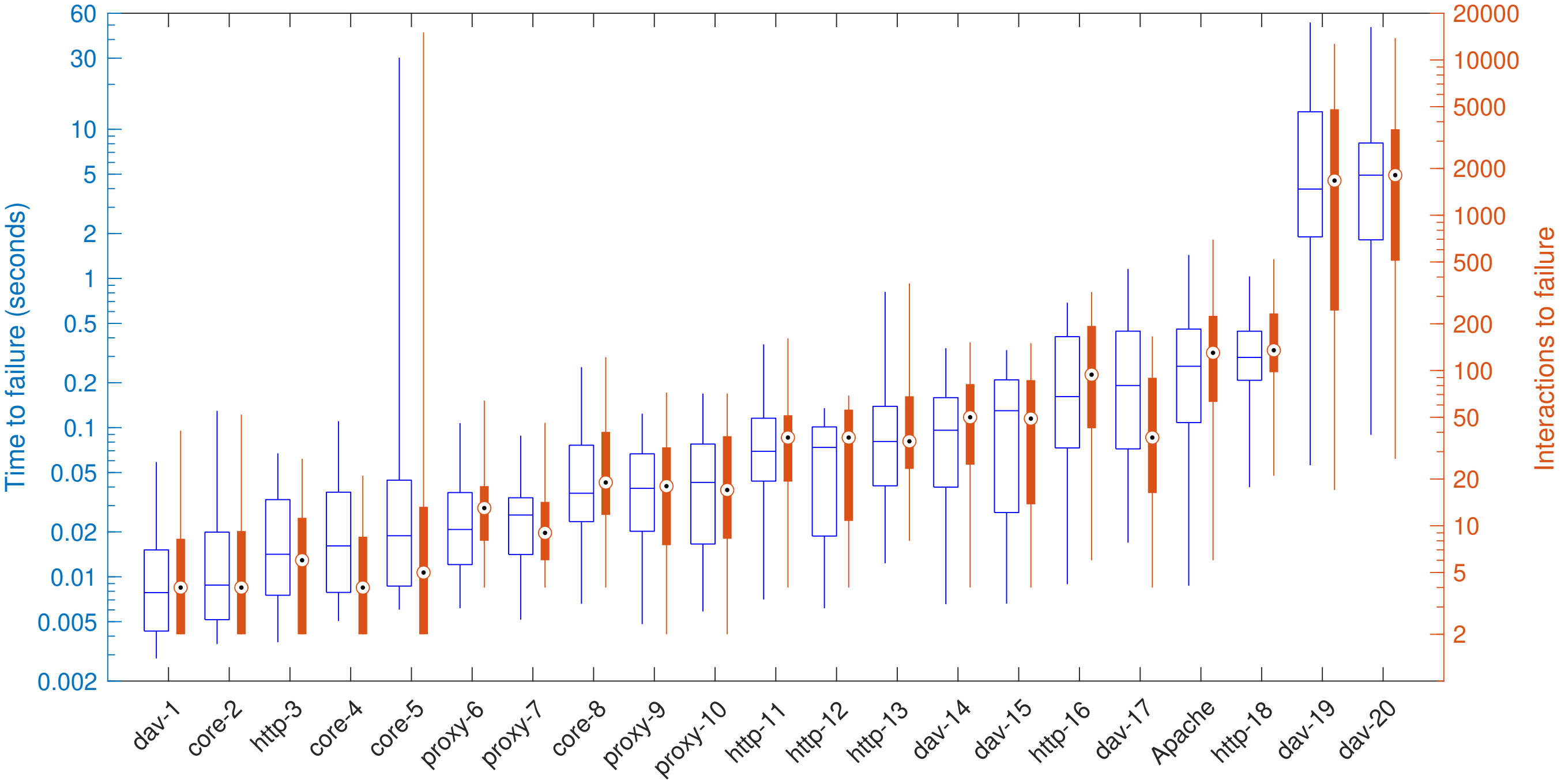

**Figure 14: Cost of detecting bug in each server/mutant. The left box with median line is the tester's execution time before rejecting the server, which includes interacting with the server and checking its responses. The right bar with median circle is the number of HTTP/1.1 messages sent and received by the tester before finding the bug. Results beyond 25%–75% are covered by whiskers.**

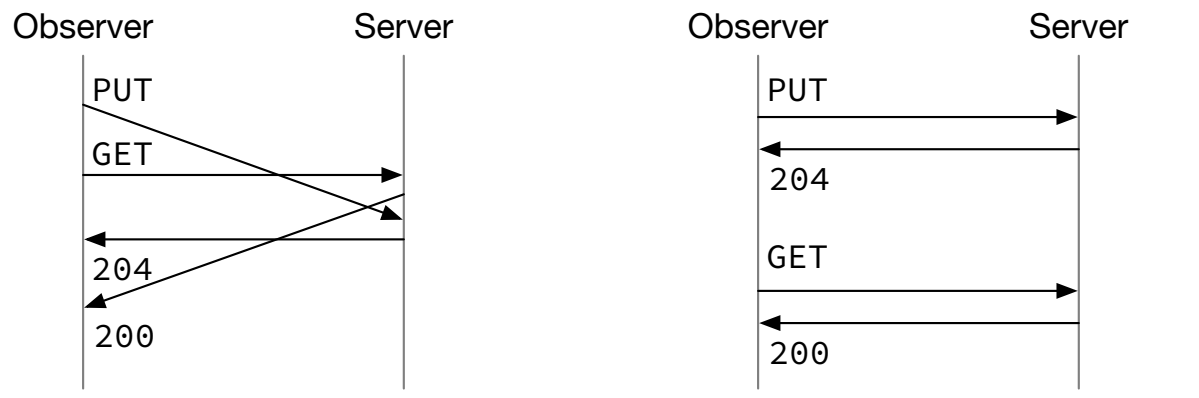

```
send (1, PUT “some” -> /key)
send (2, GET /key)
recv (1, 204 No Content)
recv (2, 200 OK: “else”)
```

Acceptable✅

```
send (1, PUT “some” -> /key)
recv (1, 204 No Content)
send (2, GET /key)
recv (2, 200 OK: “else”)
```

Unacceptable❌

**Figure 15: The trace on the left does not convince the tester that the server is buggy, because there exists a certain network delay that explains why the PUT request was not reflected in the 200 response. When the trace is ordered as shown on the right, the tester cannot imagine any network reordering that causes such observation, thus must reject the server.**

request targets within a small range, so the requests' targets are likely to be created by the tester's previous PUT requests. Narrowing the range of test case generation might improve the performance in aforementioned Mutants 18–20, but Mutant 5 shows that it could also degrade the performance of finding some bugs.

- The mutants in proxy module caused the server to forward wrong requests or responses. When the origin server part of the tester accepts a connection from the proxy, it does not know for which client the proxy is forwarding requests. Therefore, the tester needs to check the requests sent by all clients, and make sure none of them matches the incoming proxy request, before rejecting the proxy.

These examples show that the time-consuming issue of some mutants are likely caused by limitations in the test case generators. Cases like Mutant 5 can be optimized by tuning the request generator based on the tester model's runtime state, but for Mutants 18–20, the requests should be sent at specific time periods so that the resulting trace is unacceptable per specification. How to produce a specific order of messages is to be explored in future work.

## 6 RELATED WORK

### 6.1 Specifying and Testing Protocols

Modelling languages for specifying protocols can be partitioned into three styles, according to Anand et al. [1]: (1) *Process-oriented* notations that describe the SUT's behavior in a procedural style, using various domain-specific languages like our interaction trees; (2) *State-oriented* notations that specify what behavior the SUT should exhibit in a given state, which includes variants of labelled transition systems (LTS); and (3) *Scenario-oriented* notations that describe the expected behavior from an outside observer's point of view (*i.e.,* "god's-eye view").

The area of model-based testing is well-studied, diverse and difficult to navigate [1]. Here we focus on techniques that have been practiced in testing real-world programs, which includes notations (1) and (2). Notation (3) is infeasible for protocols with nontrivial nondeterminism, because the specification needs to define observer-side knowledge of the SUT's all possible internal states, making it complex to implement and hard to reason about, as shown in Figure 1.

*Process-Oriented Style: LOTOS and TorXakis.* Language of Temporal Ordering Specification (LOTOS) [6] is the ISO standard for specifying OSI protocols. It defines distributed concurrent systems as *processes* that interact via *channels*, and represents internal nondeterminism as choices among processes.

Using a formal language strongly insired by LOTOS, Tretmans and van de Laar implemented a test generation tool for symbolic transition systems called TorXakis, which has been used for testing Dropbox [21].

TorXakis provides limited support for internal nondeterminism. Unlike our testing framework that incorporates symbolic evalutation, TorXakis enumerates all possible values of internally generated data, until finding a corresponding case that matches the tester's observation. This requires the server model to generate data within a reasonably small range, and thus cannot handle generic choices like HTTP entity tags, which can be arbitrary strings.

As for network nondeterminism of interacting with multiple clients, LOTOS-style specifications requrie defining input/output channels for each client-server connection, so the tester cannot create more connections than the model's specified channels. Whereas in our application model, input and output operations are described as events, and different channels are distinguished by the events' parameter, so the model supports infinitely many connections, and the derived tester can simulate as many clients as the operating system supports.

*State-Oriented Style: NetSem, Modbat, and IOSTS.* Bishop et al. [5] have developed rigorous specifications for transport-layer protocols TCP, UDP, and the Sockets API, and validated the specifications against mainstream implementations in FreeBSD, Linux, and WinXP. Their specification represents internal nondeterminism as symbolic states of the model, which is then evaluated using a special-purpose symbolic model checker. They focused on developing a post-hoc specification that matches existing systems, and wrote a separate tool for generating test cases. Whereas, our work aims at finding bugs in server implementations, so we derived the specification into a testing program.

Using an abstract model based on extended finite state machines (EFSM), Artho and Rousset [3] have generated test cases for Java network API, which involves blocking and non-blocking communications. They have found defects in the network library `java.nio`, including unexpected exceptions and failing assertions. While their state machine specification writes assertions about *what* behavior is valid, our program model defines *how* to produce valid behavior. The difference between these validation logic was shown in Section 2 and 3.

Rusu et al. [18] have extended LTS into Input/Output Symbolic Transition System (IOSTS), and derived symbolic tests that were applied for a simple version of Common Electronic Purse Specification. In order to generate test cases, their specification IOSTS needs to be composed with a test purpose IOSTS that defines a goal of the test experiment. In comparison, our tester is derived from the specification itself, and checks whether the SUT's observable behavior is expected by the specification.

### 6.2 Reasoning about Network Delays

Koh et al. [15] have introduced "network refinement" relation, a variant of observational refinement, to specify implementations that are indistinguishable as observed over the network. Using the same specification language as ours, they have formally verified a simple "swap server" written in C, and tested the server by reordering the client-side trace to find a corresponding server-side trace that matches the server model. Whereas, we combined the server model with a network model to include all possible client-side observations.

For property-based testing against distributed applications like Dropbox, Hughes et al. [14] have introduced "conjectured events" to represent uploading and downloading events that nodes may perform at any time invisibly. We are inspired by the idea of conjecturing *when* the events might happen, and provided a more modularized language for describing nondeterminism in networked applications. By splitting the specification into server and network models, handling nondeterminism in each, we clarified *how* the servers' behavior might vary. Such modularization also makes our specification reusable, as the network model can be composed with many other server models to address network reorderings.

Besides reasoning about the packets' order of arrival, Sun et al. [20] symbolised the time elapsed to transmit packets from one end to another, and developed a symbolic-execution-based tester that found transmission-related bugs in Linux TFTP upon certain network delays. Their tester used a fixed trace of packets to interact with the server, and the generated test cases were the packets' delay time. In comparison, we specify the space of valid observations caused by possible network delays, and generate messages that trigger semantic bugs in networked applications.

## 7 FUTURE WORK

### 7.1 Test Case Generation

As discussed in Subsection 4.2, our framework allows applying heuristics to generate certain messages, but hasn't implemented configuration of the interactions' order. We hope to explore the potential of packet dynamics [20] and combine their generator's mechanism with ours. By tuning the time of sending messages, we'd expect our tester to capture bugs more effectively.

### 7.2 Shrinking

When our tester rejects a server, it reports the first failing trace as counterexample, which possibly contains irrelevant transactions. To locate the bug efficiently, we hope to *shrink* the counterexample to a minimal trace that clearly shows why the server is wrong. The existing shrinking technique for pure functional programs needs to be adjusted to our scenario, where the response of an impure server differs from one execution to another. Shrinking for interactive, nondeterministic programs deserves more exploration. A smaller counterexample might require not only fewer and smaller messages, but also reordering the messages as well.

## 7.3 Verification

Our specification language is designed for both testing and verification purposes. Using the same specification language, we have verified and tested a similar HTTP/1.1 server written in C [23]. The verified server covers a subset of HTTP/1.1 features in this paper. As we expand the verified server, we'd expect to eliminate bugs by testing it against the specification, until formal proof provides an even more rigorous correctness guarantee.

Although a limited number of test cases doesn't give us full guarantee of the server's compliance upon arbitrary inputs, we are still interested in proving that the tester is "exhaustive"—*i.e.,* for all servers that contains some bug, the tester can *eventually* generate some counterexample to reveal that bug. Having an exhaustive tester, as the number of tests increase, our confidence of the server's correctness converges to 100%.

## 7.4 Specifying and Testing HTTP/1.1 in General

In this paper, we specified a network application that covers a subset of HTTP/1.1. Whereas in real world, HTTP/1.1 is a basis for various web applications. We hope to generalize the current specification into a generic library, so that developers can specify their own web apps, and derive a tester for it.

# 8 CONCLUSION

We introduced a domain-specific language for specifying networked applications, addressing challenges of internal and network nondeterminism. We then presented a derivation framework that interprets the specification into an interactive tester program. The derived tester can reveal bugs in server implementations—including those we were unaware of—within reasonable amount of time. Our specification language is also capable of formal verification.